\documentclass[pra,twocolumn,shownopacs]{revtex4}

\usepackage{graphicx}
\usepackage{hyperref}
\usepackage{amsmath}
\usepackage{bbm}
\usepackage{graphicx}

\newcommand{\bra}[1]{\langle #1 \vert}
\newcommand{\ket}[1]{\vert #1 \rangle}
\newcommand{\Right}{\ket{\mathrm{R}}}
\newcommand{\Left}{\ket{\mathrm{L}}}

\begin{document}

\title{A quantum memory for orbital angular momentum \\photonic qubits}

\author{A. Nicolas, L. Veissier, L. Giner, E. Giacobino, D. Maxein, J. Laurat}
\email{julien.laurat@upmc.fr}
\affiliation{
Laboratoire Kastler Brossel, Universit\'{e}
Pierre et Marie Curie, Ecole Normale Sup\'{e}rieure, CNRS, 4 place
Jussieu, 75252 Paris Cedex 05, France}

\maketitle
 
\noindent Among the optical degrees of freedom, the orbital angular momentum of light \cite{Allen03} provides unique properties \cite{Torres11}, including mechanical torque action with applications for light manipulation \cite{Grier}, enhanced sensitivity in imaging techniques \cite{Sergienko13} and potential high-density information coding for optical communication systems \cite{Wang13}. Recent years have also seen a tremendous interest in exploiting orbital angular momentum at the single-photon level in quantum information technologies \cite{Mair01,Leach02}. In this endeavor, here we demonstrate the implementation of a quantum memory \cite{Lvovsky09} for quantum bits encoded in this optical degree of freedom. We generate various qubits with computer-controlled holograms, store and retrieve them on demand using the dynamic electromagnetically-induced transparency protocol. We further analyse the retrieved states by quantum tomography and thereby demonstrate fidelities exceeding the classical benchmark, confirming the quantum functioning of our storage process. Our results provide an essential capability for future networks \cite{Kimble08} exploring the promises of orbital angular momentum of photons for quantum information applications.\\

A most studied family of beams carrying orbital angular momentum (OAM) are the Laguerre-Gaussian (LG) modes, which exhibit a helical phase structure and carry an OAM that can take any integer value. Arising from the solution of the paraxial wave equation in cylindrical coordinates, these modes define an unbounded basis for transverse modes. Owing to this infinite dimensionality, the orbital angular momentum of photons raised intense theoretical and experimental efforts related to its use for encoding and processing quantum information \cite{Torres11}. 

Following the pioneering work demonstrating entanglement in this degree of freedom for photons \cite{Mair01}, great advances have been obtained in the experimental control of OAM state superpositions and their use in a variety of protocols. These advances include quantum cryptography \cite{Groblacher06}, bit commitment \cite{Langford04}, experimental quantum coin tossing \cite{Molina05}, and more recently, the demonstration of very high-dimensional entanglement \cite{Dada11,Fickler12}. Beyond their fundamental significance, these groundbreaking experiments testify the potential of the orbital angular momentum of light as quantum information carrier, holding much promise for an enhanced information coding density and processing capabilities. This promise can be extended to applications such as quantum networks, including quantum repeaters over long distances \cite{Kimble08}. For such OAM-based implementation, spatially multimode light-mater interfaces will be required.

Therefore, the ability to store OAM superpositions at the single-photon level in matter systems is of crucial importance for future developments. In recent years, significant progresses in this direction have been achieved by demonstrating the entanglement of OAM states between a photon and an ensemble of cold atoms \cite{Inoue} or the reversible mapping of bright light beams carrying OAM \cite{Pugatch07,Moretti09}. Moreover, the preservation of the handedness of the helical phase structure when storing a Laguerre-Gaussian mode has been recently demonstrated at the single-photon level \cite{Veissier2013,Ding2013}, as well as the preservation of the spatial structure for a specific mode superposition \cite{Ding2013}. However, to date no experiment has addressed the capability to store quantum bits, as achieved for instance with other degrees of freedom such as polarization in a variety of physical systems \cite{Caltech,Specht11,ICFO,GAP,Hefei}.  

In this paper, we report on the physical implementation of a quantum memory for OAM qubits. Based on a single large ensemble of cold cesium atoms and the dynamic electromagnetically-induced transparency protocol (EIT) \cite{Lvovsky09}, our device enables us to store and recall superpositions of Laguerre-Gaussian modes of opposite helicities in the single-photon regime. To access the coherence of the process, we develop an interferometer-based characterization system, including spatial mode projectors. This scheme enables us to perform a full quantum tomography of the encoded states. Fidelities after readout above 92$\%$ have been achieved, beating classical benchmark. 

\begin{figure*}[t!]
\centerline{\includegraphics[width=1.75\columnwidth]{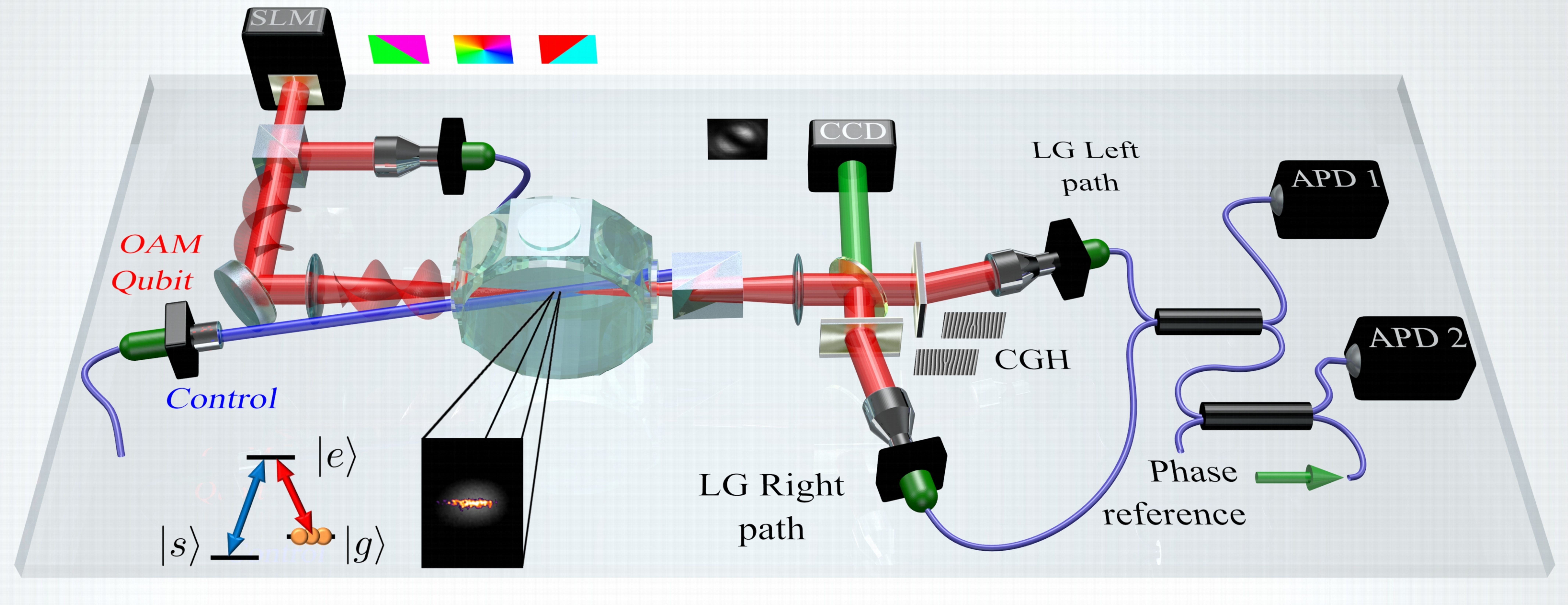}}
\caption{Experimental setup for quantum storage and analysis of OAM qubits. An orbital angular momentum photonic qubit encoded via a spatial light modulator (SLM) is coherently mapped into a large atomic ensemble of cold cesium atoms and retrieved on demand at a later time. The control and OAM qubit fields have linear orthogonal polarizations, copropagate with an angle of 1.7$^{\circ}$ and are separated after the memory interface by a Glan-Taylor polarizing beam-splitter. To fully reconstruct the density matrix of the retrieved qubits, the photons enter into a two-path interferometer, where each path includes a mode projector based on a blazed-fork computer-generated hologram (CGH) and a single-mode fiber. The two paths are arranged in a way to project the photons respectively into the $\Left$ mode (LG Left path) and the $\Right$ mode (LG Right path). Events are detected at the output of a fiber beam-splitter by single-photon counting modules (APD 1 and APD 2). The relative phase $\varphi$ between the two paths is scanned and experimentally determined by sending a phase-reference beam backward and analyzing its spatial structure at the input of the interferometer via a digital camera. The zoom in inset shows a false color image of the atomic ensemble of about 2 mm in length.}
\label{fig1}
\end{figure*}

The experimental setup is depicted in Fig. \ref{fig1}. OAM qubits are implemented with weak coherent states with an adjustable mean number of photons per pulse $\bar{n}$. The qubit basis is defined by the two Laguerre-Gaussian modes $\Right = \ket{\mathrm{LG}^{l=+1}_{p=0}}$ and $\Left = \ket{\mathrm{LG}^{l=-1}_{p=0}}$, where $p$ and $l$ are respectively the radial and azimuthal indices. These modes are the so-called doughnut modes and exhibit a vanishing intensity in the centre. We initialize the state to be stored by shaping the wavefront with a computer-controlled spatial light modulator (see Appendix B). 

The qubit is then coherently mapped into a large ensemble of cold cesium atoms in a magneto-optical trap (MOT) with an optical depth of 15 (see Appendix A). The atomic three-level $\Lambda$ system used here involves two hyperfine ground states, $|s\rangle=|6S_{1/2},F=3\rangle$ and $|g\rangle=|6S_{1/2},F=4\rangle$, and one excited state $|e\rangle=|6P_{3/2},F=4\rangle$. All the atoms are initially prepared in $|g\rangle$. While the qubit field is resonant with the $|g\rangle \leftrightarrow |e\rangle$ transition, the dynamic EIT storage requires an additional control field, resonant with $|s\rangle\leftrightarrow |e\rangle$. The control and qubit fields, which are orthogonally polarized, are locked in phase and frequency at the cesium hyperfine splitting frequency. They are copropagating in the ensemble with a 1.7$^\circ$ angle. The control power is 20 $\mu$W, with a waist of 200 $\mu$m, while the qubit field has a waist of 50 $\mu$m. 

The storage process consists in sending the qubits into the memory while the classical control beam is on, making the medium transparent and reducing the group velocity \cite{Hau}. The control beam is then ramped down to zero in 50 ns resulting in mapping the qubit into an atomic coherence. Due to the limited delay (200 ns corresponding to a light slowing by a factor $3\,.10^4$) coming from the finite optical depth and length of the sample, a fraction of the light leaks out of the sample without being stored, contributing to lowering the overall efficiency. When switching the control back on after a user-defined delay, the stored light is re-emitted into the same spatial mode due to a collective enhancement effect \cite{Liu}. 

\begin{figure}[b!]
\centerline{\includegraphics[width=0.85\columnwidth]{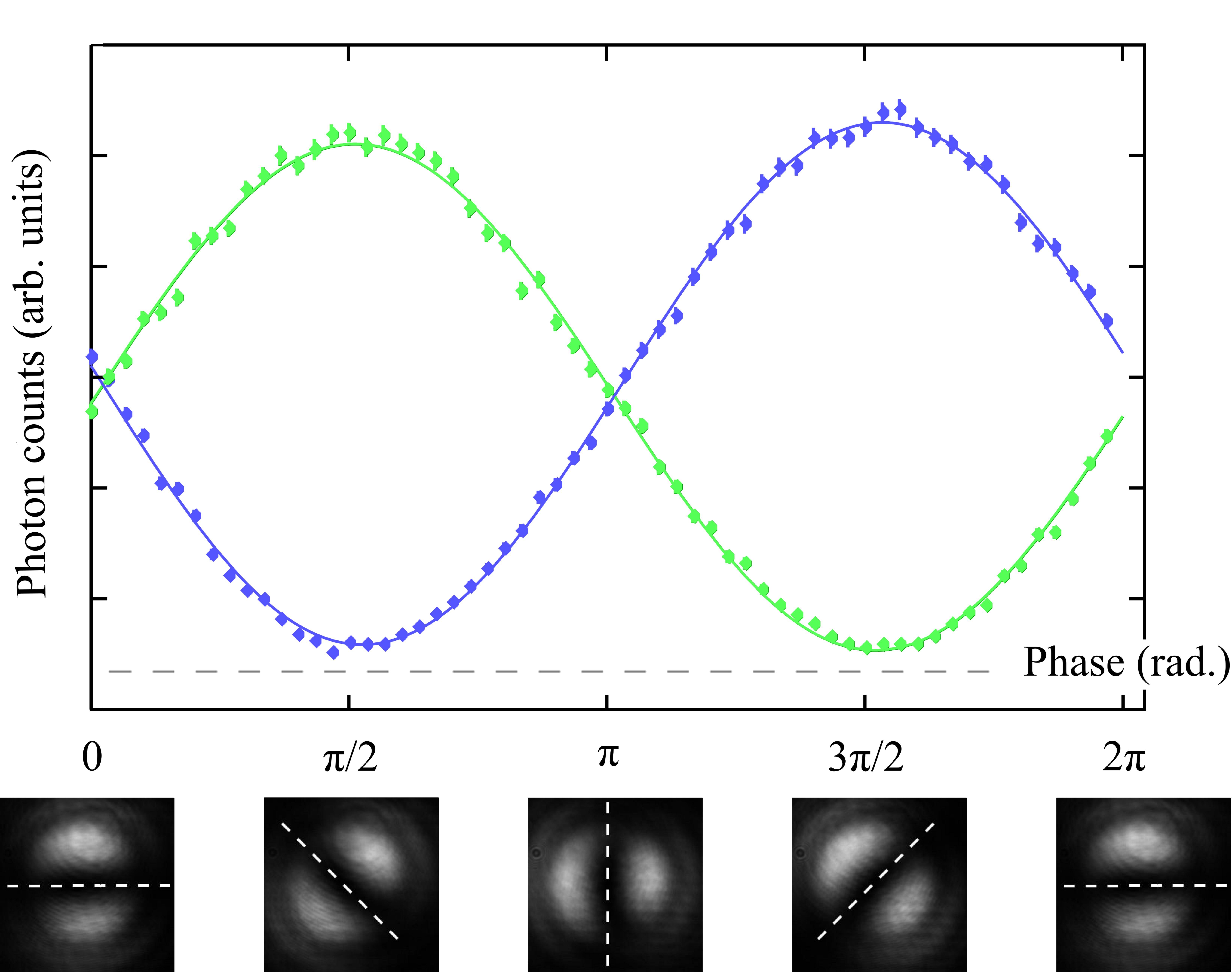}}
\caption{Experimental fringe measurements and phase analysis. Measured counts on APD 1 (dark blue) and APD 2 (light green) for a retrieved $\ket{\rm{A}}$ state when the relative phase $\varphi$ of the interferometer is scanned. No background correction has been applied and the dotted horizontal line gives the background level. The fitted visibilities are $80 \%\pm1\%$ for both fringes ($96.5 \%\pm1\%$ with correction). The displayed modes correspond to the images recorded on the  camera and used for the calibration of the relative phase (see Appendix B). The dotted white lines show the phase computed by the image analysis routine. The count rate for APD 2 has been multiplied by two in order to account for the second fibre beamsplitter used to back-propagate the phase-reference beam. Errors were estimated assuming Poissonian statistics.
}
\label{fig2}
\end{figure}

We characterize the performance of our memory device by storing a set of input qubits distributed over the Bloch sphere and by performing subsequent quantum state tomography of the retrieved states. Reconstructing the density matrix $\hat{\rho}$ of any two-dimensional state requires three linearly independent measurements, as it can be written as $\hat{\rho}=\frac{1}{2}\left(\hat{\rm{I}}+\sum\limits_{i=1}^3  S_i\,\hat{\sigma}_i\right) $ where $\hat{\sigma}_i$ are the Pauli spin operators and $\hat{\rm{I}}$ is the identity matrix. To access the $S_i$ parameters, measurements are performed in the three spatial bases defined by LG modes ($\{\Right,\Left\}$), and Hermite-Gaussian modes ($\{\ket{\rm{H}}=\left(\Right+\Left\right)/\sqrt{2}, \ket{\rm{V}}=\left(\Right-\Left\right)/\sqrt{2}\}$ and $\{\ket{\rm{D}}=\left(\Right+i\Left\right)/\sqrt{2}, \ket{\rm{A}}=\left(\Right-i\Left\right)/\sqrt{2}\}$). To this end, we develop a detection scheme based on spatial mode projectors inside an interferometer, as depicted in Fig. 1 and explained in the following. 

\begin{figure}[t!]
\centerline{\includegraphics[width=0.85\columnwidth]{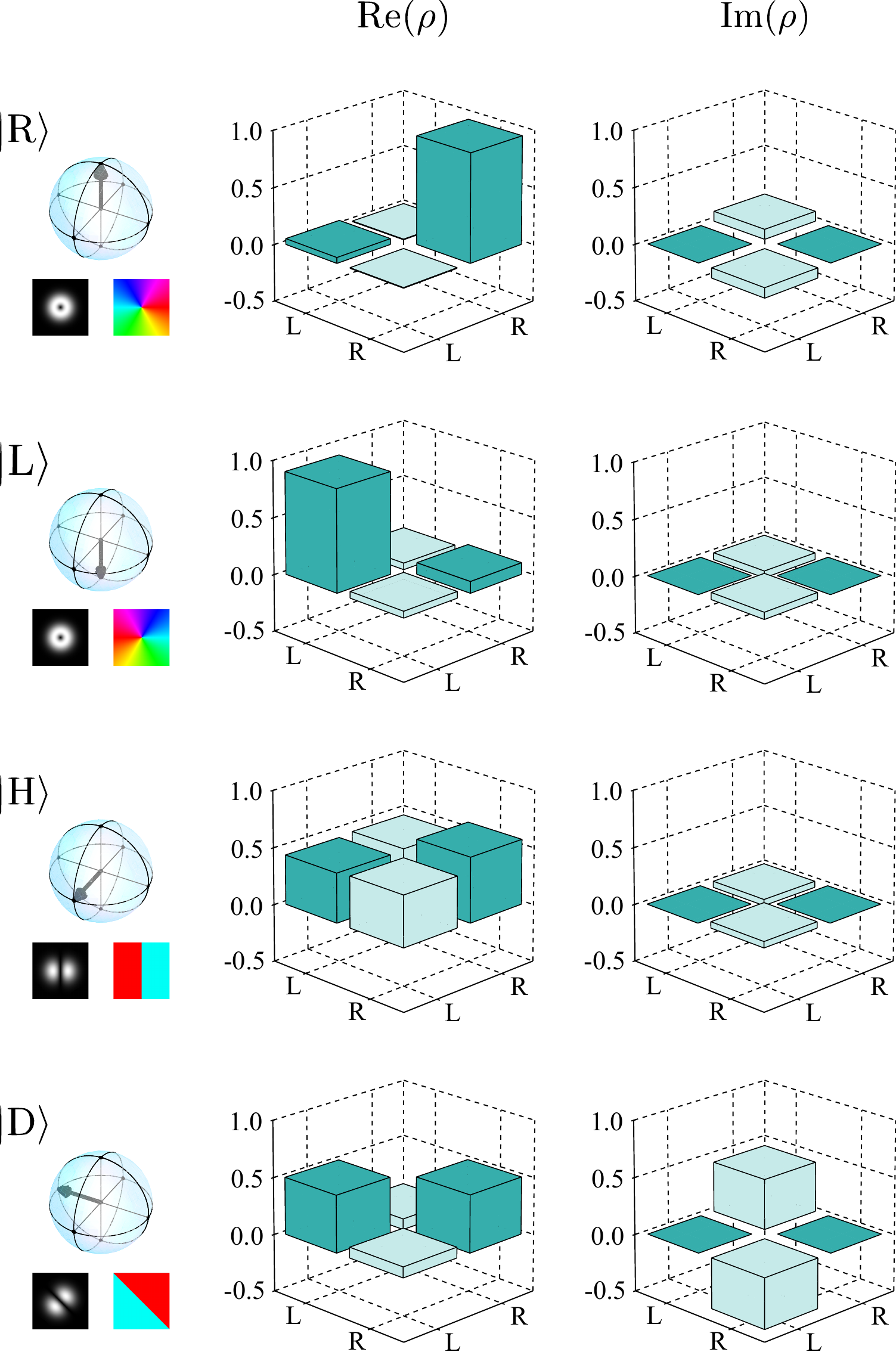}}
\caption{\textbf{Quantum tomography of the retrieved OAM qubits.} Reconstructed density matrices for the four input states $\Right$, $\Left$, $\ket{\rm{H}}=\left(\Right+\Left\right)/\sqrt{2}$ and $\ket{\rm{D}}=\left(\Right+i\Left\right)/\sqrt{2}$. The mean number of photons per pulse is here $\bar{n}=0.6$, and no background correction has been applied. The first column displays for each state its location in the Bloch sphere, the phase pattern imprinted by the SLM and the associated spatial mode.}
\label{fig3}
\end{figure}

The mode to characterize is first split on a 50/50 non-polarizing beamsplitter. Each path includes a mode projector based on a blazed fork hologram and a subsequent single-mode fibre. The light is diffracted with high efficiency ($>80\%$) into the first order \cite{Vaziri02}, resulting in the subtraction of one OAM unit in the so-called Right path and in the addition of one unit in the Left path. The resulting TEM$_{00}$ light coupled into the subsequent single-mode fibers is thus the projection of the initial mode on the $\Right$ and $\Left$ qubit states. These components are then brought to interfere via a fibre beam-splitter and the two outputs are detected by single-photon counting modules. Given the relative phase $\varphi$ of the interferometer, which is scanned and measured via a real-time image analysis (see Appendix B), this scheme gives access to the projections on the two rotated bases. Specifically, the count rates on APD 1 (respectively APD 2) at phases $\varphi = 0$, $\pi/2$, $\pi$, and $3\pi/2$ provide the qubit components along modes $\ket{\rm{V}}$, $\ket{\rm{D}}$, $\ket{\rm{H}}$ and $\ket{\rm{A}}$ (respectively $\ket{\rm{H}}$, $\ket{\rm{A}}$, $\ket{\rm{V}}$, $\ket{\rm{D}}$). Projections on the qubit basis $\{\Right,\Left\}$ are obtained by blocking alternately the two paths of the interferometer.  As an example, Figure 2 displays the interference fringes obtained on both counting modules when the phase $\varphi$ is scanned, after storage and retrieval of the $\ket{\rm{A}}$ state. The large achieved visibilities, 80\% without noise background correction and 96.5\% with correction, confirm the coherence of the storage process. 

The set of the measured relative probabilities in the three orthogonal bases enables us to reconstruct the density matrix, as well-known for polarization tomography \cite{james01}. Figure \ref{fig3} provides the density matrices for four retrieved qubits, with a storage time of 1 $\mu$s and a mean photon number equal to $\bar{n}=0.6$, well into the single-photon regime. In order to quantitatively assess the performances of the memory, we compare these states with the ideal qubit states $\ket{\Psi}$ and compute the fidelity $ \bra{\Psi} \rho \ket{\Psi} $. The values are summarized in Table \ref{fidelity}. The average fidelity is $92.5 \% \pm 2 \%$ when using the raw data, and reach $98 \% \pm 1 \%$ when the background noise is subtracted. Let us emphasize that this value is a lower bound of the achieved storage fidelity as it includes any imperfections in the mode preparation. 

\begin{table}[b!]
\begin{center}
\begin{tabular}{|l|l|r|l|}
\hline
Input mode & Raw fidelity & Corrected fidelity   \\
\hline
\raisebox{0.1ex}{$\Right$} \hspace{1.05mm}\raisebox{-1.55ex}{\includegraphics[scale=0.67]{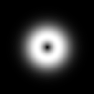}}& \raisebox{0.1ex}{$95.1 \pm 0.5 \%$} & \raisebox{0.1ex}{$99.3 \pm 0.5 \%$} \rule[-7pt]{0pt}{0.7cm} \\
\hline
\raisebox{0.1ex}{$\Left$} \hspace{1.4 mm}\raisebox{-1.55ex}{\includegraphics[scale=0.67]{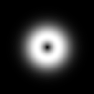}} & \raisebox{0.1ex}{$90.0 \pm 0.8 \%$} & \raisebox{0.1ex}{$97.7 \pm 0.6 \%$} \rule[-7pt]{0pt}{0.7cm}\\
\hline
\raisebox{0.1ex}{$\ket{\rm{V}}$}\hspace{1mm} \raisebox{-1.55ex}{\includegraphics[scale=0.67]{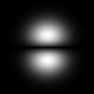}} & \raisebox{0.1ex}{$90.3 \pm 1.1 \%$} & \raisebox{0.1ex}{$98.8 \pm 0.5 \%$} \rule[-7pt]{0pt}{0.7cm}\\
\hline
\raisebox{0.1ex}{$\ket{\rm{D}}$}\hspace{1mm} \raisebox{-1.55ex}{\includegraphics[scale=0.67]{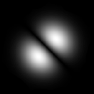}} & \raisebox{0.1ex}{$94.0 \pm 0.9 \%$} & \raisebox{0.1ex}{$98.7 \pm 0.5 \%$} \rule[-7pt]{0pt}{0.7cm}\\
\hline
\raisebox{0.1ex}{$\ket{\rm{H}}$}\hspace{1mm} \raisebox{-1.55ex}{\includegraphics[scale=0.67]{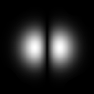}} & \raisebox{0.1ex}{$94.7 \pm 0.9 \%$} & \raisebox{0.1ex}{$98.1 \pm 0.5 \%$} \rule[-7pt]{0pt}{0.7cm}\\
\hline
\raisebox{0.1ex}{$\ket{\rm{A}}$}\hspace{1mm} \raisebox{-1.55ex}{\includegraphics[scale=0.67]{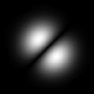}} & \raisebox{0.1ex}{$90.6 \pm 1.1 \%$} & \raisebox{0.1ex}{$96.2 \pm 0.8 \%$} \rule[-7pt]{0pt}{0.7cm}\\
\hline
\end{tabular}
\caption{\label{fidelity}\textbf{State fidelities.} Fidelities of the readout states for six input qubits without and with background noise subtraction. The mean photon number per pulse is $\bar{n}=0.6$. Errors were estimated assuming Poissonian statistics and taking into account the phase binning and residual error on the calibration of the interferometer. }
\end{center}
\end{table}

In order to conclude about the quantum character of the demonstrated storage and retrieval process, the fidelities have to be compared with the the best ones achievable using a classical memory protocol, which consists in measuring the state, storing classically the results and preparing a new state based on these data. This technique, known for instance as the intercept-resend attack in quantum cryptography, can achieve a maximum fidelity of $(N+1)/(N+2)$ for a state containing $N$ photons \cite{Popescu95}, leading to the well-known $2/3$ limit for a single-photon. In our case, as the experiment is performed with weak coherent states, the Poissonian statistics has also to be taken into account, resulting in a weighted average depending on the mean-photon number per pulse $\bar{n}$. Additionally, it can be shown that the non-unity efficiency $\eta$ of the storage and retrieval process can also increase the maximally achievable classical fidelity. By explicitly taking into account these two parameters,  Ref. \cite{Specht11} provided a fidelity threshold (see also \cite{ICFO}), showing that indeed the quantum character of a memory can be assessed with weak coherent states. This threshold approaches unity when $\bar{n}$ is increasing, thereby requiring critically to test the device with coherent pulses with a low mean photon number. 

\begin{figure}[t!]
\centerline{\includegraphics[width=0.85\columnwidth]{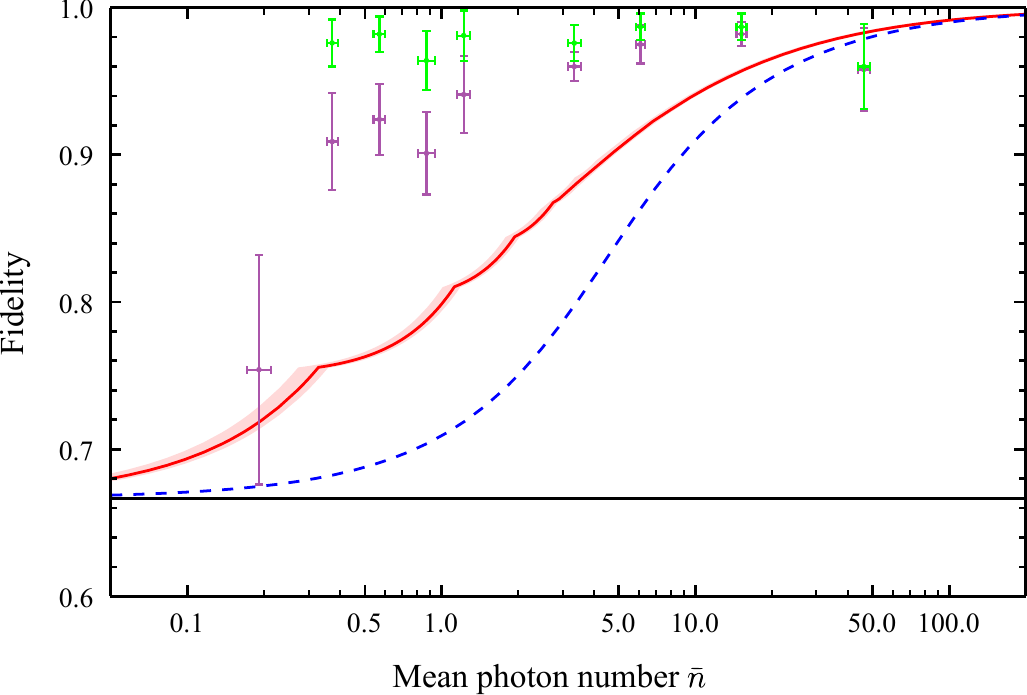}}
\caption{\textbf{Average fidelities of the retrieved qubits and quantum storage.} The state fidelity, averaged over the six input qubits, is given as a function of the mean photon-number per pulse $\bar{n}$. The purple points correspond to the raw data while the green ones are corrected from background noise. The blue dotted line gives the classical limit for a memory with unity storage and readout efficiency and the red line shows the classical limit for the actual efficiency of our memory device (the pink shaded area represents the error bar on the efficiency). Vertical and horizontal error bars indicate the standard deviations of fidelities and mean photon numbers for the six input states, respectively.}
\label{fig4}
\end{figure}

Figure \ref{fig4} provides the achieved average fidelities (raw and background-corrected) as a function of the mean photon number $\bar{n}$, together with the aforementioned classical limits. The pink shaded region corresponds to the boundary taking into account the measured storage and retrieval efficiency $\eta=15\pm2$\%. Our results beat the classical limit by several standard deviations in a large range of mean photon numbers, confirming the quantum character of our device. For very small value of $\bar{n}$, the fidelity drops due to the background noise.

The memory time of our MOT-based system is currently 15 $\mu$s, as limited by inhomogeneous Zeeman broadening due to residual magnetic field and motional dephasing resulting from the angle between the control and signal fields. Further cooling of the atoms and optical pumping into $|F=4,m_F=0\rangle$ would readily improve the memory time around 100 $\mu$s, as limited by the loss of atoms from the excitation region. To achieve larger values, as required for practical quantum networking applications \cite{Sangouard2011}, other trapping techniques have to be implemented, enabling optical storage in the second range as demonstrated for instance in light-shift compensated optical lattices \cite{Dudin2013}. OAM qubit storage in rare earth ion doped crystals \cite{Lvovsky09} would also be of large interest, with potential long storage time and without the possible issue of the transverse size limitation in an optical lattice.

In summary, we have experimentally demonstrated the operation of a faithful quantum memory for qubits encoded in the orbital angular momentum degree of freedom and based on a single atomic ensemble. No dual rail implementation with spatially separated ensembles is required, as usually done for the polarization degree of freedom, owing here to the intrinsic spatially multimode nature of the storing medium. We have shown that our device is operating better than any classical counterpart, providing a unique tool for investigating quantum networking protocols involving this encoding that holds much promise for an increase in information coding density. Given the parameters of our atomic ensemble and the scaling of LG beam size in $\sqrt{l}$, the capacity of our spatial multimode memory can be estimated around 100 qubit modes \cite{Grodecka2012}. Future developments of the work presented here include the demonstration of the reversible mapping of OAM entanglement between remote units and the extension of the storage capability to higher OAM values, whose manipulation can highly benefit from recent developments in OAM sorting techniques.

\acknowledgements
We thank A. Zeilinger and R. Fickler for providing fork holograms, and M.J. Padgett and D. Tasca for their assistance with the SLM. We also thank M. Scherman and S. Burks for their valuable contributions in the early stage of the experiment. This work is supported by the ERA-Net CHIST-ERA (QScale), the ERA-Net.RUS (Nanoquint), the Institut Francilien de Recherche sur les Atomes Froids (IFRAF) and by the European Research Council (ERC) Starting grant HybridNet. A.N. acknowledges the support from the Direction G\'en\'erale de l'Armement (DGA). J.L. is a member of the Institut Universitaire de France.

\appendix
\section{Experimental sequence}
The experiment is conducted in sequences, at a repetition rate of 66 Hz, each of them including a stage for the atomic cloud to build up (11.5 ms for MOT loading and 650 $\mu$s for further cooling by optical molasses) and a stage for memory operations (3 ms). In order to avoid inhomogeneous Zeeman broadening, the magnetic field gradient is switched off after the MOT loading and residual magnetic fields are measured via a microwave spectroscopy and compensated down to 5 mG. The memory trials (writing, storage and retrieval) are repeated 200 times per MOT cycle, with a period of 5 $\mu$s. Photons are detected by avalanche photodiodes (SPCM-AQR-14-FC), and recorded with a time resolution of 10 ns.

\section{OAM qubit generation}
The optical qubits are implemented with weak coherent states at the single-photon level, temporally shaped by an acousto-optic modulator to a half-gaussian profile with a typical width of 300 ns. The phase structure is imprinted by reflection on a computer-controlled pure phase spatial light modulator (Hamamatsu LCOS-SLM X10468-02) with a resolution of 792$\times$600 pixels. Examples of phase pattern settings are given in Fig.~1 in false colors. The $\Right$ and $\Left$ modes (i.e. Laguerre-Gaussian modes) are generated with a rotating phase pattern and the different superpositions (i.e. Hermite-Gaussian modes) with a $\pi$ phase jump pattern.

\section{Accessing the interferometer phase} The interferometer phase $\varphi$ is measured by sending backward a phase-reference beam (nW level). The light is detected on a digital camera at the second output of the first beam-splitter, as sketched in Fig. \ref{fig1}. On one path, one unit of OAM is added while on the other path one unit is subtracted. The recombination of the two resulting Laguerre-Gaussian modes with opposite helicities leads to a Hermite-Gaussian mode and its orientation is given by half the relative phase $\varphi$. Images are recorded at a rate of $8$ Hz, thus averaging over a few MOT cycles. A single computer triggers the experiment, records the APD events and the images. It associates a timestamp to both events and images. Using this timestamp and a Python-written routine to automatically detect the orientation of the images on the camera, a relative phase $\varphi$ is associated to each event.  The relative phase is discretized into 60 bins of 6$^\circ$ each. In order to avoid light going back to the avalanche photodiodes, the reference beam is sent only during the MOT building stage when the APDs are gated off.


\begin{thebibliography}{99}

\bibitem{Allen03} L. Allen, S.M. Barnett and M.J. Padgett (eds) Optical angular momentum. (IOP, 2003).

\bibitem{Torres11} J.P. Torres and L. Torner, Twisted photons: applications of light with orbital angular momentum (Wiley-VCH, 2011).

\bibitem{Grier} D.G. Grier,  A revolution in optical manipulation, Nature \textbf{424}, 810-816 (2003).

\bibitem{Sergienko13} N. Uribe-Patarroyo, A. Fraine, D.S. Simon, O. Minaeva and A.V. Sergienko, Object identification using correlated orbital angular momentum states, Phys. Rev. Lett. \textbf{110}, 043601 (2013).

\bibitem{Wang13} J. Wang \textit{et al.}, Terabit free-space data transmission employing orbital angular momentum multiplexing, Nature Photon. \textbf{6}, 488-496 (2012).

\bibitem{Mair01} A. Mair, A. Vaziri, G. Weihs and A. Zeilinger, Entanglement of the orbital angular momentum states of photons, Nature \textbf{412}, 313-316 (2001).

\bibitem{Leach02} J. Leach, M.J. Padgett, S.M. Barnett, S. Franke-Arnold and J. Courtial, Measuring the orbital angular momentum of a single photon, Phys. Rev. Lett. \textbf{88}, 257901 (2002).

\bibitem{Lvovsky09} A.I. Lvovsky, B. Sanders and W. Tittel, Optical quantum memory, Nature Photon. \textbf{3}, 706-714 (2009).

\bibitem{Kimble08} H.J. Kimble, The quantum internet, Nature \textbf{453}, 1023-1030 (2008). 

\bibitem{Groblacher06} S. Groblacher, T. Jennewein, A. Vaziri, G. Weihs and A. Zeilinger, Experimental quantum cryptography with qutrit, New. J. Phys. \textbf{8}, 75 (2006).

\bibitem{Langford04} N.K. Langford \textit{et al.}, Measuring entangled qutrits and their use for quantum bit commitment, Phys. Rev. Lett. \textbf{93}, 053601 (2004).

\bibitem{Molina05} G. Molina-Terriza, A. Vaziri, R. Ursin and A. Zeilinger, Experimental quantum coin tossing, Phys. Rev. Lett. \textbf{94}, 040501 (2005).

\bibitem{Dada11} A.C. Dada, J. Leach, G.S. Buller, M.J. Padgett and E. Andersson, Experimental high-dimensional two-photon entanglement and violations of generalized Bell inequalities, Nature Phys. \textbf{7}, 677-680 (2011).

\bibitem{Fickler12} R. Fickler \textit{et al.}, Quantum entanglement of high angular momenta, Science \textbf{338}, 640-643 (2012).

\bibitem{Inoue} R. Inoue \textit{et al.}, Entanglement of orbital angular momentum states between an ensemble of cold atoms and a photon, Phys. Rev. A \textbf{74}, 053809 (2006).

\bibitem{Pugatch07} R. Pugatch, M. Shuker, O. Firstenberg, A. Ron and N. Davidson, Topological stability of optical vortices, Phys. Rev. Lett. \textbf{98}, 203601 (2007).

\bibitem{Moretti09} D. Moretti, D. Felinto and J.W.R. Tabosa, Collapses and revivals of stored orbital angular momentum of light in a cold-atom ensemble, Phys. Rev. A \textbf{79}, 023825 (2009).

\bibitem{Veissier2013} L. Veissier \textit{et al.}, Reversible optical memory for twisted photons, Opt. Lett. \textbf{38}, 712-714 (2013).

\bibitem{Ding2013} D.-S. Ding, Z.-Y. Zhou, B.S. Shi, and G-C. Guo, Single-photon level quantum image memory based on cold atomic ensembles, Nature Commun. \textbf{4}, 2527 (2013).

\bibitem{Caltech} K.S. Choi, H. Deng, J. Laurat and H.J. Kimble, Mapping photonic entanglement into and out of a quantum memory, Nature \textbf{452}, 67-72 (2008).

\bibitem{Specht11} H.P. Specht \textit{et al.}, A single-atom quantum memory, Nature 473, 190-193 (2011).

\bibitem{ICFO} M. G\"undogan, P. Ledingham, A. Almasi, M. Cristiani and H. de Riedmatten, Quantum storage of a photonic polarization qubit in a solid, Phys. Rev. Lett. \textbf{108}, 190504 (2012). 

\bibitem{GAP} C. Clausen, F. Bussi\`eres, M. Afzelius and N. Gisin, Quantum storage of heralded polarization qubits in birefringent and anisotropically absorbing materials, Phys. Rev. Lett. \textbf{108}, 190503 (2012).

\bibitem{Hefei} Z-Q. Zhou, W-B. Lin, M. Yang, C-F. Li and G-C. Guo, Realization of reliable solid-state quantum memory for photonic polarization qubit, Phys. Rev. Lett. \textbf{108}, 190505 (2012).

\bibitem{Hau} L.V. Hau, S.E. Harris, Z. Dutton and C.H. Behroozi, Light speed reduction to 17 metres per second in an ultracold atomic gas, Nature 397, 594-598 (1999).

\bibitem{Liu} C. Liu, Z. Dutton, C.H. Behroozi and L.V. Hau, Observation of coherent optical information storage in an atomic medium using halted light pulses, Nature \textbf{409}, 490-493 (2001).

\bibitem{Vaziri02} A. Vaziri, G. Weihs and A. Zeilinger, Superpositions of the orbital angular momentum for applications in quantum experiments, J. Opt. B \textbf{4}, S47 (2002).

\bibitem{james01} D.F.V. James, P.G. Kwiat, W.J. Munro and A.G. White, Measurement of qubits, Phys. Rev. A \textbf{64}, 052312 (2001).

\bibitem{Popescu95} S. Massar and S. Popescu, Optimal extraction of information from finite quantum ensembles, Phys. Rev. Lett. \textbf{74}, 1259 (1995).

\bibitem{Sangouard2011} N. Sangouard, C. Simon, H. de Riedmatten and N. Gisin, Quantum repeaters based on atomic ensembles and linear optics, Rev. Mod. Phys. \textbf{83}, 33-80 (2011).

\bibitem{Dudin2013} Y.O. Dudin, L. Li and A. Kuzmich, Light storage on the time scale of a minute, Phys. Rev. A \textbf{87}, 031801(R) (2013).

\bibitem{Grodecka2012} A. Grodecka-Grad, E. Zeuthen and A.S. S{\o}rensen, High-capacity spatial multimode quantum memories based on atomic ensembles, Phys. Rev. Lett. \textbf{109}, 133601 (2012).

\end{thebibliography}
\end{document}